\documentclass[twocolumn]{aastex62}
\usepackage{IEEEtrantools}
\bstctlcite{IEEEexample:BSTcontrol}
\usepackage{natbib}
\usepackage{amsmath}
\usepackage{mathtools}
\usepackage{rotating}
\usepackage{lipsum}

\usepackage{hyperref}
\usepackage{url}

\received{October 9, 2020}
\revised{January 5, 2021}
\accepted{January 7, 2021}

\shorttitle{Metallicities from Extinction Curves}
\shortauthors{Hassan Fathivavsari}

\begin{document}

\title{\bf{Constraining the Metallicities of Damped Ly$\alpha$ Systems Using Extinction Curves}}

\correspondingauthor{Hassan Fathivavsari}
\email{h.fathie@gmail.com}

\author{Hassan Fathivavsari}
\affiliation{School of Astronomy, Institute for Research in Fundamental Sciences (IPM), P. O. Box 19395-5531, Tehran, Iran}

\begin{abstract}

In this paper, we present a new method to constrain the metallicities of high redshift damped Ly\,$\alpha$ (DLA) absorbers using observed extinction curves. This is the first time such an approach is employed to constrain the metallicities of extragalactic absorbers. To demonstrate our method, we use the spectra of 13 quasars and one GRB with DLA absorbers detected along their sightlines. By using the Kramers-Kronig (KK) relation, which relates the wavelength-integrated extinction to the total volume occupied by dust per hydrogen nucleon, we set some robust lower limits on the metallicity of the DLAs. The resulting lower limits are all consistent with the DLA metallicities from the literature. The GRB extinction curve exhibits a very strong 2175\,\textup{\AA} extinction bump. We try to constrain the metallicity of the GRB DLA by modeling the GRB extinction curve using dust models with two (graphite and silicates) and three (PAH, hydrogenated amorphous carbon, and silicates) dust components. The two-component model resulted in a metallicity of $Z\,\sim$\,$-$0.45 while the three-component model gives $Z\,\sim$\,$-$0.50. On the other hand, the lower limit from the KK approach for this DLA is $Z\,\ge$\,$-$0.60. Modeling a large sample of extinction curves with 2175\,\textup{\AA} extinction bump and measured DLA metallicities would allow a thorough comparison between the KK and the model-dependent approach. In cases where the precise measurement of the metallicity of a DLA is not possible (e.g. due to the saturation of important absorption lines), the approach presented in this paper can be used to constrain the metallicity.

\end{abstract}


\section{Introduction} \label{sec:intro}


The interstellar medium (ISM) of the Galaxy is composed of gas and dust. The gas-phase abundances of the elements present in the ISM can be measured from the absorption lines they produce in the spectra of background stars \citep{1975ARA&A..13..133S,1992ApJ...401..706S,1995ApJ...445..196S,1997ApJ...489..672W,2012ApJS..199....8G}. On the other hand, the dust-phase abundance of an element is derived by subtracting off its gas-phase abundance from a reference abundance \citep{1994ApJ...430..650S,1998ApJ...493..583V,2010A&A...517A..45V,2013A&A...560A..88D}. The solar abundance \citep{1989GeCoA..53..197A,2009ARA&A..47..481A} is usually taken as the reference abundance. However, the protosolar abundances augmented by Galactic chemical enrichment and the abundances of the unevolved early B stars are also adopted as the reference abundances \citep{2015ApJ...809..120M,2017ApJ...850..138M}.

The presence of dust in the ISM also attenuates and extincts the light of background stars through absorption and scattering. The extinction (which is the sum of absorption and scattering) is wavelength-dependent in such a way that photons with bluer wavelengths tend to be more extincted. Due to this tendency, the wavelength-dependence of extinction is often referred to as \emph{reddening}. The amount of dust responsible for the observed extinction can be estimated using the Kramers-Kronig (KK) relation \citep{1969ApJ...158..433P}. As shown in \citet{2005ApJ...622..965L}, the KK relation relates the wavelength-integrated extinction to the total volume (per hydrogen nucleon) occupied by dust through

\begin{equation} \label{eq:1}
\int_{0}^{\infty} \frac{A_{\lambda}}{N_{\rm H}} d\lambda = 1.086 \times 3 \pi^{2} F \frac{V_{\rm dust}}{\rm H},
\end{equation}

\noindent
where $N_{\rm H}$ is the hydrogen column density, and $F$ is a dimensionless factor and is a function of the grain shape and the static (zero-frequency) dielectric constant $\epsilon_{0}$ of the dust grain material. The value of $F$ for different dust materials can be calculated using the equation\,4 in \citet{2015ApJ...809..120M}. Since $A_{\lambda}$ is only known for a limited range of wavelengths, in practice, the integral in equation\,\ref{eq:1} is usually calculated over the wavelength range of 912\,\textup{\AA}\,$\le$\,$\lambda$\,$\le$\,1000\,$\mu$m, and the result is considered as a lower limit. For the diffuse ISM, the integral of $A_{\lambda}$/$N_{\rm H}$ over this wavelength range is 1.49\,$\times$\,10$^{-25}$\,mag\,cm$^{-3}$\,H$^{-1}$. In principle, this contains the contributions from different species of dust materials such as silicates and carbonaceous grains.

It has been well recognized that silicate is a major component of the ISM dust \citep{1994ApJ...422..164K,2001ApJ...548..296W,2001ApJ...550L.213L,2001ApJ...554..778L,2011MNRAS.410.1932Z,2015ApJ...809..120M,2015ApJ...811...38W,2017ApJ...850..138M,2020P&SS..18304627G,2020MNRAS.497.2190M}. The presence of silicate grains in the ISM is primarily revealed through the detection of the 9.7\,$\mu$m Si$-$O stretching and 18\,$\mu$m O$-$Si$-$O bending absorption features \citep{1994A&A...292..641J,2010ARA&A..48...21H}. The smooth and featureless structures of these absorption features indicate that the silicate grains in the ISM are predominantly of amorphous composition \citep{2001ApJ...554..778L,2004ApJ...609..826K,2008ApJ...685.1046L}.

\begin{figure}
\centering
\begin{tabular}{c}
\includegraphics[width=0.99\hsize]{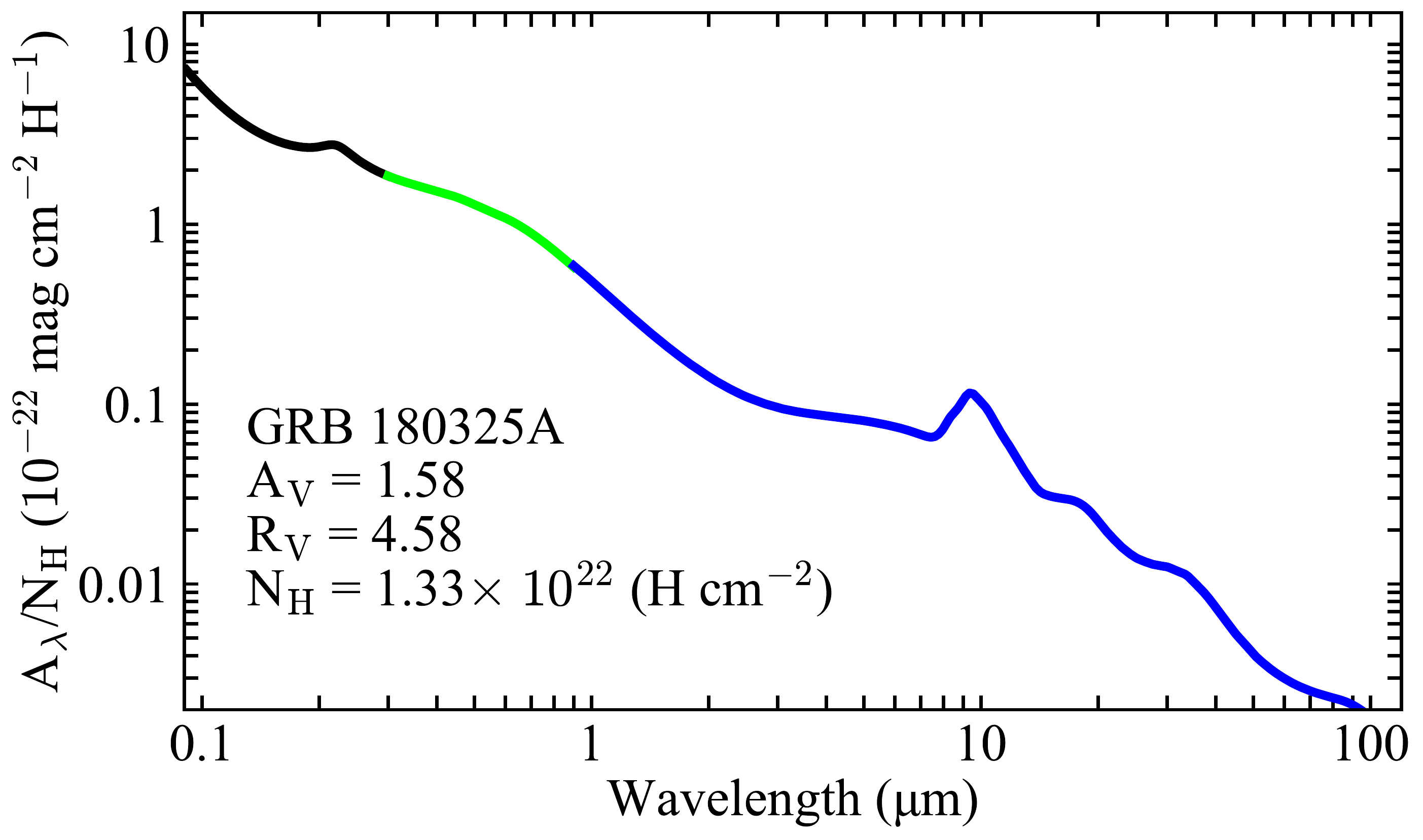}
\end{tabular}
\caption{The extinction curve from the far UV to the far IR for GRB\,180325A, with the \citet{1990ApJS...72..163F} curve for the 3.3\,$<$\,$\lambda^{-1}$\,$<$\,11\,$\mu$m$^{-1}$ (black line), the \citet{2007ApJ...663..320F} curve for the 1.1\,$<$\,$\lambda^{-1}$\,$<$\,3.3\,$\mu$m$^{-1}$ (green line), and the \citet{2015ApJ...811...38W} curve for the $\lambda$\,$>$\,0.9\,$\mu$m (blue line).}
 \label{AvNH}
\end{figure}

Are silicate grains alone sufficient to explain the observed extinction\,? If we assume that all Si, Mg, and Fe elements of solar abundances \citep{2009ARA&A..47..481A} are locked up in silicate grains with a stoichiometric composition of MgFeSiO$_{4}$, the upper limit that we get for the total silicate volume per H nucleon is 
\begin{equation} \label{eq:2}
V_{\rm sil}/{\rm H} \approx  (M_{\rm sil}/M_{\rm H}) / \rho_{\rm sil} \times m_{\rm H} \approx 2.71 \times 10^{-27}\,{\rm cm}^{3}\,{\rm H}^{-1},
\end{equation}

\noindent
where $M_{\rm sil}/M_{\rm H}$ is the silicate to hydrogen mass, $\rho_{\rm sil}$ is the mass density of silicate, and $m_{\rm H}$\,=\,1.66\,$\times$\,10$^{-26}$\,g is the mass of hydrogen atom. With F\,$\approx$\,0.7 \citep[see][]{2015ApJ...809..120M}, the wavelength-integrated extinction originating from the silicate component of the dust is at most $\sim$\,6.01\,$\times$\,10$^{-26}$\,mag\,cm$^{3}$\,H$^{-1}$. This is not sufficient to explain the observed lower limit of 1.49\,$\times$\,10$^{-25}$\,mag\,cm$^{-3}$\,H$^{-1}$. The inability of silicate dust in fully explaining the observed extinction implies that there must be other dust components present in the ISM, with carbonaceous dust being the most favored one.

The presence of carbonaceous dust in the ISM is spectroscopically revealed through the 3.4\,$\mu$m aliphatic C$-$H stretching absorption feature \citep{1990MNRAS.243..400A,2002ApJS..138...75P}, and the \emph{unidentified} infrared emission bands at 3.3, 6.2, 7.7, 8.6, 11.3 and 12.7\,$\mu$m \citep{1984A&A...137L...5L,1985ApJ...290L..25A}. The 2175\,\textup{\AA} extinction bump is also widely attributed to the presence of small graphite grains \citep{1965ApJ...142.1681S,1993ApJ...414..632D,2009MNRAS.394.2175P}, or polycyclic aromatic hydrocarbon (PAH) molecules \citep{2001ApJ...554..778L,2010ApJ...712L..16S,2011ApJ...733...91X,2013ApJS..207....7M,2015ApJ...810...39B}. The bump, which is the strongest absorption feature observed in the ISM, is thought to arise from the $\pi$$-$$\pi^{*}$ transition in $sp^{2}$ hybridization of aromatic carbon.  As outlined above, the carbonaceous dust must account for more than half of the total observed extinction. Indeed, all modern extinction models assume that the observed extinction is predominantly caused by the combination of carbonaceous and silicate grains of various sizes \citep{1984ApJ...285...89D,1993ApJ...402..441L,1994ApJ...422..164K,2001ApJ...548..296W,2001ApJ...550L.213L,2001ApJ...554..778L,2009ApJ...690L..56L,2010ApJ...710..648L,2015ApJ...811...38W,2015ApJ...809..120M,2017ApJ...850..138M,2020MNRAS.497.2190M}.

In this paper, we introduce a method to derive some lower limits for the metallicity of high redshift DLAs using the observed extinction curves. We demonstrate our method by applying it on the spectra of 13 quasars and one GRB along which some DLA absorbers are identified. Here, we take two different approaches: the first approach, which is independent of any dust model, is to use the KK relation of \citet{1969ApJ...158..433P}, and the second approach is to model the observed extinction curve in terms of two- and three-component dust models. These techniques can be used as a complement to the metal absorption line technique especially in cases where precise metallicity is not available. This is best suited for absorbers associated with GRBs, due to their simple spectral energy distributions which lack prominent emission lines \citep{2011A&A...532A.143Z,2012ApJ...753...82Z,2018ApJ...860L..21Z,2018MNRAS.479.1542Z}, and quasar absorbers with $z_{\rm DLA}$\,$\sim$\,$z_{\rm QSO}$ such as eclipsing and ghostly DLAs \citep{2015MNRAS.454..876F,2016MNRAS.461.1816F,2017MNRAS.466L..58F,2018MNRAS.477.5625F,2020ApJ...888...85F,2020ApJ...901..123F}. In latter systems, the similarity of the absorption and emission redshift reduces the complications that may arise due to the presence of quasars intrinsic emission lines.

\begin{figure*}
\centering
\begin{tabular}{c}
\includegraphics[width=0.90\hsize]{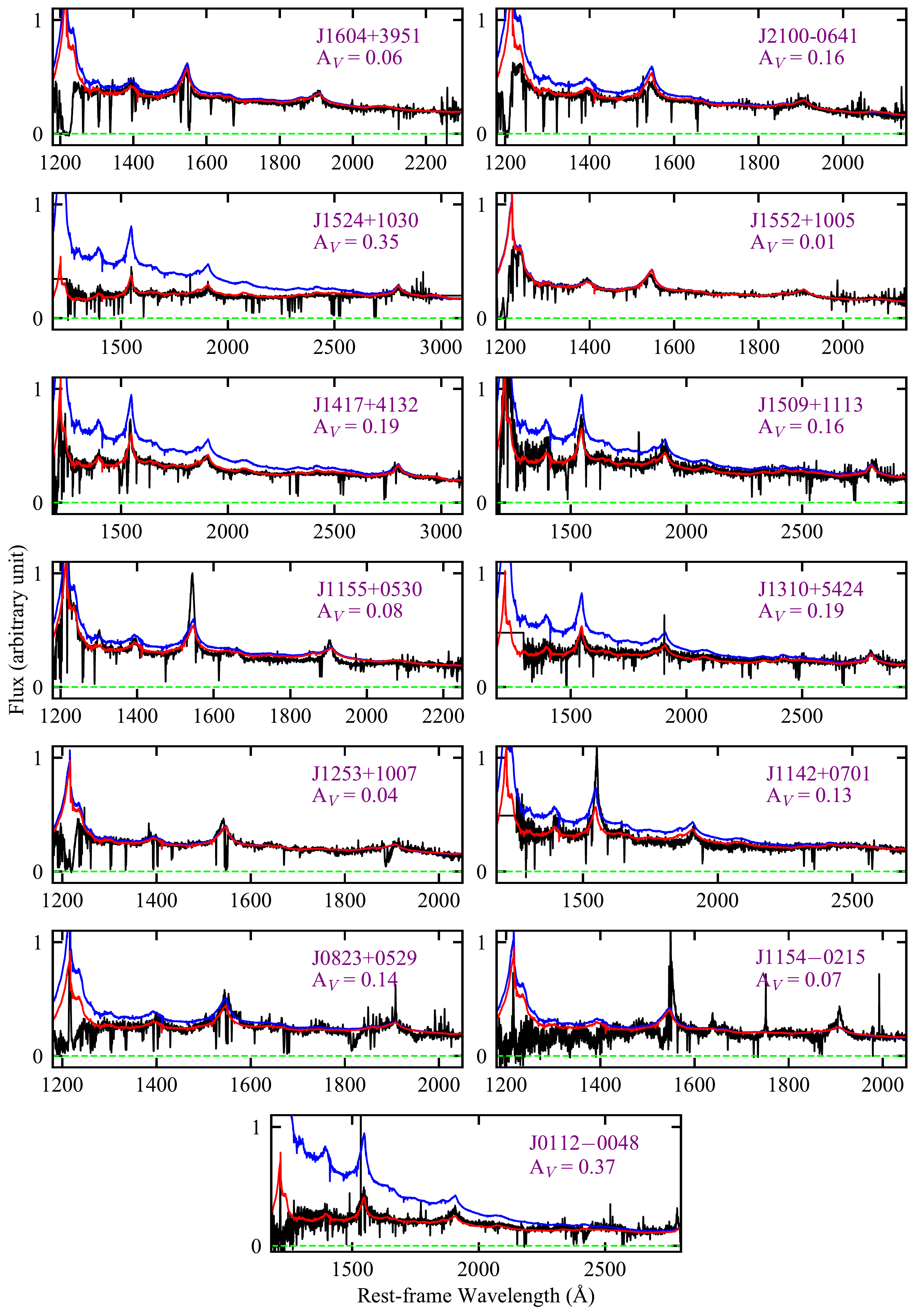}
\end{tabular}
\caption{Spectra of quasars in our sample. In each panel, the blue curve shows the quasar template spectrum of \citet{2016A&A...585A..87S}, and the red curve shows the same quasar template spectrum but reddened by the SMC extinction law.}
 \label{extplots}
\end{figure*}

\section{The Data}

In this paper, we use the spectra of 13 quasars and one GRB to demonstrate our method. These quasars are selected for two reasons: 1) they all have a DLA at $z_{\rm abs}$\,$\sim$\,$z_{\rm em}$, and 2) the metallicities of their DLAs, which are measured from high resolution spectra, are available in the literature. Table\,\ref{table1} summarizes the quasar redshift ($z_{\rm em}$), the DLA redshift ($z_{\rm abs}$), the neutral hydrogen column density of the DLA ($N$(H\,{\sc i})), along with the references from which they are compiled. The similarity of the DLA absorption and quasar emission redshifts allows a more accurate extraction of the extinction curve from the quasar spectra. None of these quasars exhibits the 2175\,\textup{\AA} extinction bump. However, a strong 2175\,\textup{\AA} bump is detected in the GRB spectrum. The bump is detected at the redshift of a DLA absorber with $z_{\rm abs}$\,=\,2.2486 and neutral hydrogen column density of log\,$N$(H\,{\sc i}/cm$^{-2}$)\,=\,22.30\,$\pm$\,0.14 \citep{2018ApJ...860L..21Z}. This bump is the only unambiguous detection of such a feature in a GRB spectrum in more than a decade. \citet{2018ApJ...860L..21Z} derived a lower limit of [Zn/H]\,$\ge$\,$-$0.98 for the metallicity of the DLA.

Since our method is based on exploiting the observed extinction curves to constrain DLA metallicities, we first explain how we construct the extinction over wavelength from 912\,\textup{\AA} to 1\,cm. In the case of the GRB\,180325A, for the wavelength range 3.3\,$<$\,$\lambda^{-1}$\,$<$\,11\,$\mu$m$^{-1}$, we represent the extinction by the \citet{1990ApJS...72..163F} extinction law with the extinction parameters taken  from \citet{2018ApJ...860L..21Z}: $c_{1}$\,=\,$-$1.95\,$\mu$m, $c_{2}$\,=\,1.28, $c_{3}$\,=\,2.92, $c_{4}$\,=\,0.52, $\gamma$\,=\,1.16\,$\mu$m$^{-1}$, $x_{0}$\,=\,4.538\,$\mu$m$^{-1}$, $R_{V}$\,=\,4.58, and $A_{V}$\,=\,1.58. For the wavelength range 1.1\,$<$\,$\lambda^{-1}$\,$<$\,3.3\,$\mu$m$^{-1}$ the extinction curve is defined by a cubic spline interpolation between a set of optical/IR anchor points and a pair of UV anchor points \citep{2007ApJ...663..320F}. For 0.9\,$\mu$m\,$<$\,$\lambda$\,$<$\,1\,cm, the extinction is approximated by the model extinction calculated from the standard silicate-graphite-PAH model of \citet{2015ApJ...811...38W}. The final extinction curve is shown in Fig.\,\ref{AvNH}. For the ease of illustration, we only show in Fig.\,\ref{AvNH} the extinction curve up to 100\,$\mu$m.

In the case of the quasars, we employ the template-matching technique to measure the extinction, $A_{\rm V}$, and construct the extinction curve for the wavelength range 3.3\,$<$\,$\lambda^{-1}$\,$<$\,11\,$\mu$m$^{-1}$ \citep{2007ApJ...663..320F}. Here, we iteratively redden the template quasar spectrum of \citet{2016A&A...585A..87S} using the Small Magellanic Cloud (SMC) extinction curve \citep{2003ApJ...594..279G} until the reddened template spectrum best matches the observation \citep[for more details see][]{2020ApJ...888...85F}. The results are shown in Fig.\,\ref{extplots}. In this figure, the blue and red curves are the template quasar spectrum before and after applying the reddening, respectively. Moreover, for the wavelength range 1.1\,$<$\,$\lambda^{-1}$\,$<$\,3.3\,$\mu$m$^{-1}$ and 0.9\,$\mu$m\,$<$\,$\lambda$\,$<$\,1\,cm, the extinction curve is approximated following the same approach as was employed to the GRB (see above).

\section{Results}

\subsection{Metallicity Inferred from the Kramers-Kronig Relation}

In this section, we employ the KK relation of \citet{1969ApJ...158..433P} to set a robust lower limit on the metallicity of the DLAs. For this purpose, we first define a grid of metellicities ranging from $-$2.0 to 0.0 with the step of 0.05\,dex. For each metallicity in the grid, we calculate the area under the {\it expected} extinction curve. In practice, we calculate V$_{\rm dust}$/H and then use equation\,\ref{eq:1} to convert it to the area under the extinction curve. The metallicity at which the expected area is equal to the observed one is taken as the lower limit of the metallicity. The area under the observed extinction curves  are listed in the 8th column of Table\,\ref{table1}. Since these areas are calculated from the spectral range of 912\,\textup{\AA} to 1\,cm, they are considered as lower limits.

For the dust composition, we take into account three different compositions: 1) graphite and Fe-bearing silicates (i.e., Olivine MgFeSiO$_{4}$). For this composition, the term $F\,V_{\rm dust}/{\rm H}$ in equation\,\ref{eq:1} is defined as

\begin{equation} \label{eq:3}
F\, \frac{V_{\rm dust}}{{\rm H}} =  F_{\rm gra}\,\frac{V_{\rm gra}}{{\rm H}}\,+\,F_{\rm sil}\,\frac{V_{\rm sil}}{{\rm H}}.
\end{equation}

\noindent
2) graphite, Fe-lacking silicates (i.e., enstatite MgSiO$_{3}$), and pure iron \citep{2015ApJ...801..110P,2016ApJ...825..136D}. Here, the term $F\,V_{\rm dust}/{\rm H}$ is defined as

\begin{equation} \label{eq:4}
F\, \frac{V_{\rm dust}}{{\rm H}} =  F_{\rm gra}\,\frac{V_{\rm gra}}{{\rm H}}\,+\,F_{\rm sil}\,\frac{V_{\rm sil}}{{\rm H}}\,+\,F_{\rm Fe}\,\frac{V_{\rm Fe}}{{\rm H}}.
\end{equation}

\noindent
3) graphite, Fe-lacking silicates, and iron oxides (i.e., FeO, Fe$_{2}$O$_{3}$, and Fe$_{3}$O$_{4}$). For this composition, the term $F\,V_{\rm dust}/{\rm H}$ is defined as

\begin{equation} \label{eq:5}
F\, \frac{V_{\rm dust}}{{\rm H}} =  F_{\rm gra}\,\frac{V_{\rm gra}}{{\rm H}}\,+\,F_{\rm sil}\,\frac{V_{\rm sil}}{{\rm H}}\,+\,F_{\rm Fe}^{\rm oxide}\, \frac{V_{\rm Fe}^{\rm oxide}}{{\rm H}},
\end{equation}

\noindent
where

\begin{equation} \label{eq:6}
F_{\rm Fe}^{\rm oxide}\, \frac{V_{\rm Fe}^{\rm oxide}}{{\rm H}} =  \,F_{\rm FeO}\,\frac{V_{\rm FeO}}{{\rm H}}\,+\,F_{\rm Fe_{2}O_{3}}\,\frac{V_{\rm Fe_{2}O_{3}}}{{\rm H}}\,+\,F_{\rm Fe_{3}O_{4}}\,\frac{V_{\rm Fe_{3}O_{4}}}{{\rm H}}.
\end{equation}

\noindent
Here, we assume that all Si and Fe atoms, and 70 percent of carbon atoms are locked up in dust. For the iron oxides, we also assume that Fe atoms are equally shared among the three different iron oxides \citep{2020arXiv201109440Z}. The $F$ factors for graphite, silicates, iron, and iron oxides are listed in Table\,8 of \citet{2020arXiv201109440Z}.

\begin{figure*}
\centering
\begin{tabular}{c}
\includegraphics[width=0.89\hsize]{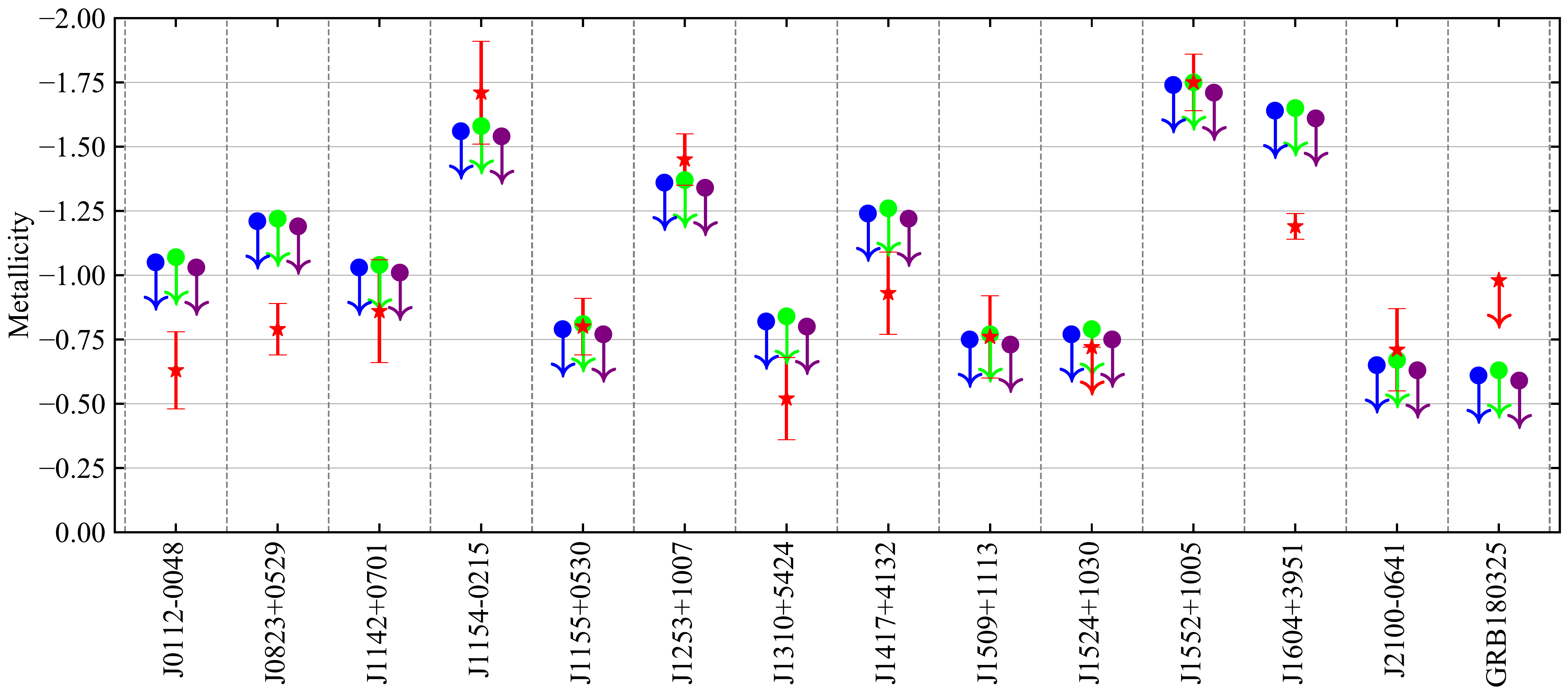}
\end{tabular}
\caption{Visual representation of the observed and the KK approach metallicities. Here, the red symbols are the observed data points, and the blue, green, and purple symbols represent the lower limits from the KK approach.}
 \label{Zplots}
\end{figure*}

Table\,\ref{table1} summarizes the results of our KK approach. For better visualization of the results, we also plot in Fig\,\ref{Zplots} the observed (i.e. [X/H]) and the KK lower limit (i.e. $Z_{\rm KK}$) metallicities of the DLAs. As shown in this figure, we achieve $Z_{\rm KK}$\,$\le$\,[X/H] for all DLAs along the quasars line of sight, within the uncertainty of the observed metallicities. For the GRB line of sight, our new lower limit is $\sim$\,0.4\,dex higher than the one reported by \citet{2018ApJ...860L..21Z}. This implies that the DLA along the GRB\,180325A could have a metallicity even higher that $\sim\,-$0.6.

\begin{figure}
\centering
\begin{tabular}{c}
\includegraphics[width=0.99\hsize]{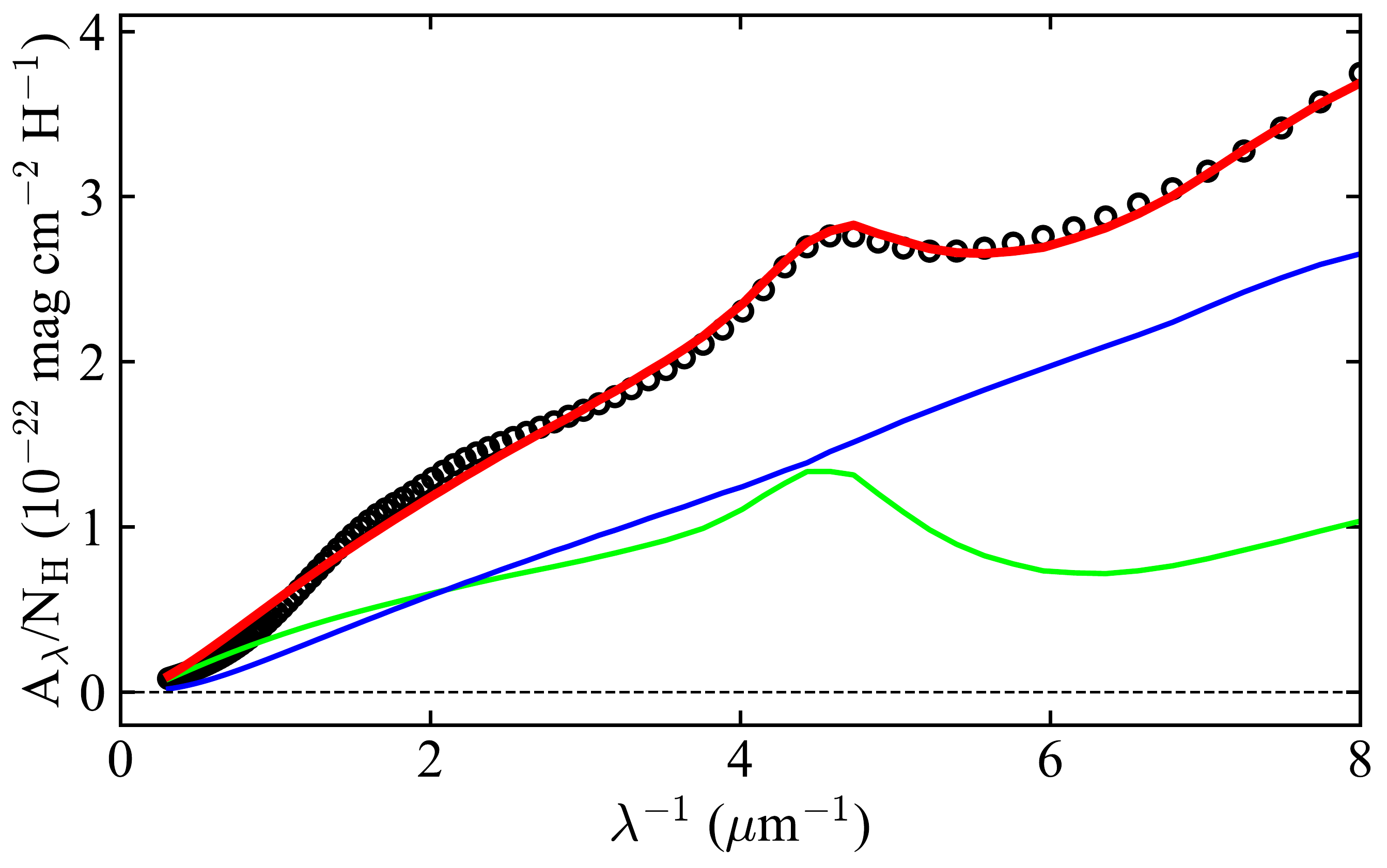}
\end{tabular}
\caption{Observed (black circles) and model (red line) extinction curves of GRB\,180325A for the two-component dust model. The red line is the combination of the contributions from the silicate (blue line) and graphite (green line) dust grains.}
 \label{modelfit}
\end{figure}

\begin{table*}
\caption{Results of the KK approach. {\it Col.\,1}: the name of the quasar and/or GRB. {\it Col.\,2}: the emission redshift of the quasar. {\it Col.\,3}: the absorption redshift of the DLA. {\it Col.\,4}: logarithm of the neutral hydrogen column density of the DLA.  {\it Col.\,5}: the observed metallicity of the DLA. {\it Col.\,6}: the low-ionization metal transition used in measuring the observed metallicity of the DLA.  {\it Col.\,7}: the observed visual extinction. {\it Col.\,8}: the area under the observed extinction curve in unit of 10$^{-27}$\,mag\,cm$^{3}$\,H$^{-1}$. {\it Col.\,9}: The KK lower limit of the metallicity by considering graphite and Fe-bearing silicates as being the main components of the dust. {\it Col.\,10}: the KK lower limit of the metallicity by considering graphite, Fe-lacking silicates, and iron oxides as being the main components of the dust. {\it Col.\,11}: the KK lower limit of the metallicity by considering graphite, Fe-lacking silicates, and pure iron as being the main components of the dust. {\it Col.\,12}: the references from which $z_{\rm em}$, $z_{\rm abs}$, $N$(H\,{\sc i}), and [X/H] are compiled. Here (1), (2), (3), and (4) refer to \citet{2016MNRAS.461.1816F}, \citet{2015PASP..127..167B}, \citet{2016MNRAS.463.3021B}, and \citet{2018ApJ...860L..21Z}, respectively.}
\centering 
 \setlength{\tabcolsep}{6.0pt}
\renewcommand{\arraystretch}{1.05}
\begin{tabular}{c c c c c c c c c c c c}
\hline\hline 
ID  & $z_{\rm em}$ & $z_{\rm abs}$ & $N_{\rm HI}$ &  [X/H]  &    X    &  $A_{V}$  &  $A_{\rm int}$  &  $Z_{\rm KK}^{1}$  &  $Z_{\rm KK}^{2}$  &  $Z_{\rm KK}^{3}$   & ref.    \\
\hline 
J0112$-$0048 & 2.1485 & 2.1493 & 21.95\,$\pm$\,0.10  &  $-$0.63\,$\pm$\,0.15  &   Si\,{\sc ii}  &  0.37  &  11.6  &   $\ge-$1.05  &  $\ge-$1.07  &  $\ge-$1.03   &  (1)    \\
J0823+0529    & 3.1875 & 3.1910 & 21.70\,$\pm$\,0.10  &  $-$0.79\,$\pm$\,0.10  &   Si\,{\sc ii}  &  0.14  &  8.1   &   $\ge-$1.21  &  $\ge-$1.22  &  $\ge-$1.19   &  (1)    \\
J1142+0701    & 1.8700 & 1.8407 & 21.50\,$\pm$\,0.15  &  $-$0.86\,$\pm$\,0.20  &   Si\,{\sc ii}  &  0.13  &  12.2 &   $\ge-$1.03  &  $\ge-$1.04  &  $\ge-$1.01   &  (2)    \\
J1154$-$0215 & 2.1810 & 2.1853 & 21.75\,$\pm$\,0.10  &  $-$1.71\,$\pm$\,0.20  &   Si\,{\sc ii}  &  0.07  &  3.6   &   $\ge-$1.56  &  $\ge-$1.58  &  $\ge-$1.54   &  (1)    \\
J1155+0530    & 3.4800 & 3.3260 & 21.05\,$\pm$\,0.10  &  $-$0.80\,$\pm$\,0.11  &   S\,{\sc ii}   &  0.08  &  21.2 &   $\ge-$0.79  &  $\ge-$0.81  &  $\ge-$0.77   &  (2)    \\
J1253+1007    & 3.0150 & 3.0312 & 21.30\,$\pm$\,0.10  &  $-$1.45\,$\pm$\,0.10  &   Si\,{\sc ii}  &  0.04  &  5.7   &   $\ge-$1.36  &  $\ge-$1.37  &  $\ge-$1.34   &  (1)    \\
J1310+5424    & 1.9300 & 1.8006 & 21.45\,$\pm$\,0.15  &  $-$0.52\,$\pm$\,0.16  &   Si\,{\sc ii}   &  0.19  &  19.6 &   $\ge-$0.82  &  $\ge-$0.84  &  $\ge-$0.80   &  (2)    \\
J1417+4132    & 2.0200 & 1.9509 & 21.85\,$\pm$\,0.15  &  $-$0.93\,$\pm$\,0.16  &   Zn\,{\sc ii}  &  0.19  &  7.5   &   $\ge-$1.24  &  $\ge-$1.26  &  $\ge-$1.22   &  (2)    \\
J1509+1113    & 2.1100 & 2.0283 & 21.30\,$\pm$\,0.15  &  $-$0.76\,$\pm$\,0.16  &   S\,{\sc ii}   &  0.16  &  23.3   &   $\ge-$0.75  &  $\ge-$0.77  &  $\ge-$0.73   &  (2)    \\
J1524+1030    & 2.0600 & 1.9409 & 21.65\,$\pm$\,0.15  &  $\ge-$0.72  &   Zn\,{\sc ii}   &  0.35  &  22.0   &   $\ge-$0.77  &  $\ge-$0.79  &  $\ge-$0.75   &  (2)    \\
J1552+1005    & 3.7220 & 3.6010 & 21.10\,$\pm$\,0.10  &  $-$1.75\,$\pm$\,0.11  &   S\,{\sc ii}   &  0.01  &  2.4   &   $\ge-$1.74  &  $\ge-$1.75  &  $\ge-$1.71   &  (3)    \\
J1604+3951    & 3.1542 & 3.1633 & 21.75\,$\pm$\,0.10  &  $-$1.19\,$\pm$\,0.05  &   S\,{\sc ii}   &  0.06  &  3.0   &   $\ge-$1.64  &  $\ge-$1.65  &  $\ge-$1.61   &  (2)    \\
J2100$-$0641 & 3.1295 & 3.0924 & 21.05\,$\pm$\,0.15  &  $-$0.71\,$\pm$\,0.16  &   S\,{\sc ii}   &  0.11  &  29.0   &   $\ge-$0.65  &  $\ge-$0.67  &  $\ge-$0.63   &  (2)    \\
GRB\,180325A & .... & 2.2486 & 22.30\,$\pm$\,0.14  &  $\ge-$0.98  &   Zn\,{\sc ii}   &  1.58  &  32.1   &   $\ge-$0.61  &  $\ge-$0.63  &  $\ge-$0.59   &  (4)    \\
\hline 
\end{tabular}
\label{table1} 
\end{table*}

\subsection{Metallicity Inferred from Modeling the Observed Extinction Curve} 

As mentioned before, the GRB extinction curve exhibits a strong 2175\,\textup{\AA} extinction bump, which is a common characteristic feature of Galactic sightlines. In this section, we try to estimate some lower limits for the metallicity of the DLA detected along the GRB line of sight by modeling the GRB extinction curve. For this purpose, we consider two kinds of dust models: 1) the two-component model of \citet{1977ApJ...217..425M} and 2) the three-component model of \citet{2011A&A...525A.103C}.

\subsubsection{Two-Component Dust Model} \label{sect:two-comp}

In this section, we model the observed extinction curve using an extinction model with two dust components: amorphous silicate and graphite \citep{1977ApJ...217..425M,1984ApJ...285...89D,2001ApJ...548..296W,2015ApJ...811...38W}. Here, we show that the adopted metallicity is very important in reproducing the observed extinction curve. This would allow us to put a constraint on the metallicity of the DLA. Similar to \citet{2015ApJ...809..120M,2017ApJ...850..138M}, an exponentially cutoff power-law size distribution is adopted for both dust components:

\begin{equation} \label{eq:7}
\frac{1}{n_{\rm H}}\frac{dn_{i}}{da} = B_{i} a^{-\alpha_{i}} \,{\rm exp}(-a/a_{c,i})
\end{equation}

\noindent
Here $a$ is the radius of the dust particles, which is in the range of 50\,\textup{\AA}\,$<$\,$a$\,$<$\,2.5\,$\mu$m. The dust particles are assumed to be spherical so that the Mie theory can be used to compute extinction cross-sections \citep{1977ApJ...217..425M}. $n_{\rm H}$ is the number density of hydrogen nuclei, $dn_{i}$ is the number density of each type of dust ('$i$' represents either silicate or graphite) with radii between $a$ and $a\,+\,da$, $\alpha_{i}$ and $a_{c,i}$ are, respectively, the power law index and exponential cutoff size, and $B_{i}$ relates to the total amount of each dust type present in the cloud. 

The total extinction per hydrogen column at wavelength $\lambda$ is given by

\begin{equation} \label{eq:8}
A_{\lambda}/N_{\rm H} = 1.086 \sum_{i} \int da \frac{1}{n_{\rm H}}\frac{dn_{i}}{da} C_{{\rm ext},i}(a, \lambda),
\end{equation}

\noindent
where the summation is over the silicate and graphite dust grains, $N_{\rm H}$ is the hydrogen column density, and $C_{{\rm ext},i}(a, \lambda)$ is the extinction cross-section of dust grains of size $a$ at wavelength $\lambda$. For each silicate and graphite dust grain, $C_{{\rm ext},i}(a, \lambda)$ can be calculated from the Mie theory by using the dielectric functions of 'astronomical' silicate and graphite from \citet{1984ApJ...285...89D}.

We use the Levenberg-Marquardt method \citep{1992nrfa.book.....P} to fit the observed extinction curve from 0.1 to 8\,$\mu$m$^{-1}$. We evaluate the extinction at 100 wavelengths, equally spaced in ln\,$\lambda$, and minimize $\chi^{2}$\,=\,$\chi^{2}_{1}$\,+\,$\chi^{2}_{2}$, which is the sum of two error functions:

\begin{equation} \label{eq:9}
\chi^{2}_{1} = \sum_{i} \frac{[A_{\rm obs}(\lambda_{i})/10^{21} - A_{\rm mod}(\lambda_{i})/10^{21}]^{2}}{\sigma^{2}_{i}},
\end{equation}

\noindent
and

\begin{equation} \label{eq:10}
\chi^{2}_{2} = 0.4[{\rm max}(\tilde{\rm C}, 1) - 1]^{1.5} + 0.4[{\rm max}(\tilde{\rm Si}, 1) - 1]^{1.5},
\end{equation}

\noindent
where $A_{\rm obs}(\lambda_{i})$ is the observed extinction,  $A_{\rm mod}(\lambda_{i})$ is the extinction computed by the model, $\sigma_{i}$ is the weight, $\tilde{\rm C}$\,=\,[C/H]$_{\rm dust}$/[C/H]$_{\rm DLA}$, and $\tilde{\rm Si}$\,=\,[Si/H]$_{\rm dust}$/[Si/H]$_{\rm DLA}$. Here, [C/H]$_{\rm DLA}$ and [Si/H]$_{\rm DLA}$ represent the number of C and Si atoms relative to hydrogen for each adopted metallicity. On the other hand, [C/H]$_{\rm dust}$ and [Si/H]$_{\rm dust}$ are calculated using the following equations:

\begin{multline} \label{eq:11}
[{\rm C/H}]_{\rm dust} = (B_{\rm C}/\mu_{\rm C}\,m_{\rm H}) \\
\times \int da (4\,\pi/3) a^{3} \rho_{\rm gra} a^{-\alpha_{\rm C}} {\rm exp}(-a/a_{c,{\rm C}})
\end{multline}

\begin{multline} \label{eq:12}
[{\rm Si/H}]_{\rm dust} = (B_{\rm S}/\mu_{\rm sil}\,m_{\rm H}) \\
\times \int da (4\,\pi/3) a^{3} \rho_{\rm sil} a^{-\alpha_{\rm S}} {\rm exp}(-a/a_{c,{\rm S}})
\end{multline}

\noindent
where $\mu_{\rm C}$\,(=\,12) is the atomic weight of carbon, $\mu_{\rm sil}$\,(=\,172) is the atomic weight of silicate grains with the stoichiometric composition of MgFeSiO$_{4}$, $\rho_{\rm gra}$ is the mass density of graphite, and $B_{\rm C}$ and $B_{\rm S}$ are, respectively, related to the total amount of graphite and silicate present in the cloud. We take $\sigma_{i}$\,=\,1\,$\sigma_{\rm obs}$ for 2\,$\mu$m$^{-1}$\,$<$\,$\lambda^{-1}$\,$<$\,8\,$\mu$m$^{-1}$ and $\sigma_{i}$\,=\,5\,$\sigma_{\rm obs}$ for $\lambda^{-1}$\,$<$\,2\,$\mu$m$^{-1}$ since the observed extinction in the infrared region is uncertain. Here, $\sigma_{\rm obs}$\,=\,0.2 is the standard deviation of the observed data points estimated from the mean signal-to-noise ratio (SNR) of the observed spectrum (i.e. SNR\,$\sim$\,5). $\chi^{2}_{2}$ in equation\,\ref{eq:10} is a \emph{penalty} which keeps the model-consumed C and Si abundances from grossly exceeding the maximum values allowed by the assumed metallicity.

Similar to the previous section, we assume that all Si and 70 percent of C atoms are locked up in dust grains. We then fit the observed extinction curve and reach $\chi^{2}$\,=\,1.5 for the metallicity of $Z$\,$\sim$\,$-$0.45. Figure\,\ref{modelfit} shows the resulting fit, and Table\,\ref{table2} lists the parameters of the fit. In Fig.\,\ref{modelfit}, the black circles show the observed extinction curve and the red curve shows the model extinction curve, which is the combination of silicate (blue curve) and graphite (green curve) contributions. We also checked the effect of increasing $N^{\rm C}_{\rm dust}$/$N^{\rm C}_{\rm tot}$, and found that this would lead to a slightly lower metallicity, i.e. $-$0.45\,$<$\,$Z$\,$<$\,$-$0.50, but with higher $\chi^{2}$ ($\sim$\,2.6). Moreover, we also found that smaller $N^{\rm C}_{\rm dust}$/$N^{\rm C}_{\rm tot}$ ratio would result in a slightly higher metallicity ($Z$\,$\sim$\,$-$0.40).

\begin{table}
\caption{Best model parameters for fitting the GRB extinction curve with a two-component dust model.} 
\centering 
 \setlength{\tabcolsep}{12.8pt}
\renewcommand{\arraystretch}{1.05}
\begin{tabular}{c c c c}
\hline\hline 
ID & $\alpha_{i}$ & $a_{c,i}$($\mu$m) & $B_{i}$ \\
\hline 
Silicate & 3.21 & 0.22 & 14.0\,$\times$\,10$^{-25}$  \\
Graphite & 3.27 & 0.37 & 2.6\,$\times$\,10$^{-25}$  \\
\hline 
\end{tabular}
\label{table2} 
\end{table}

\begin{figure}
\centering
\begin{tabular}{c}
\includegraphics[width=0.99\hsize]{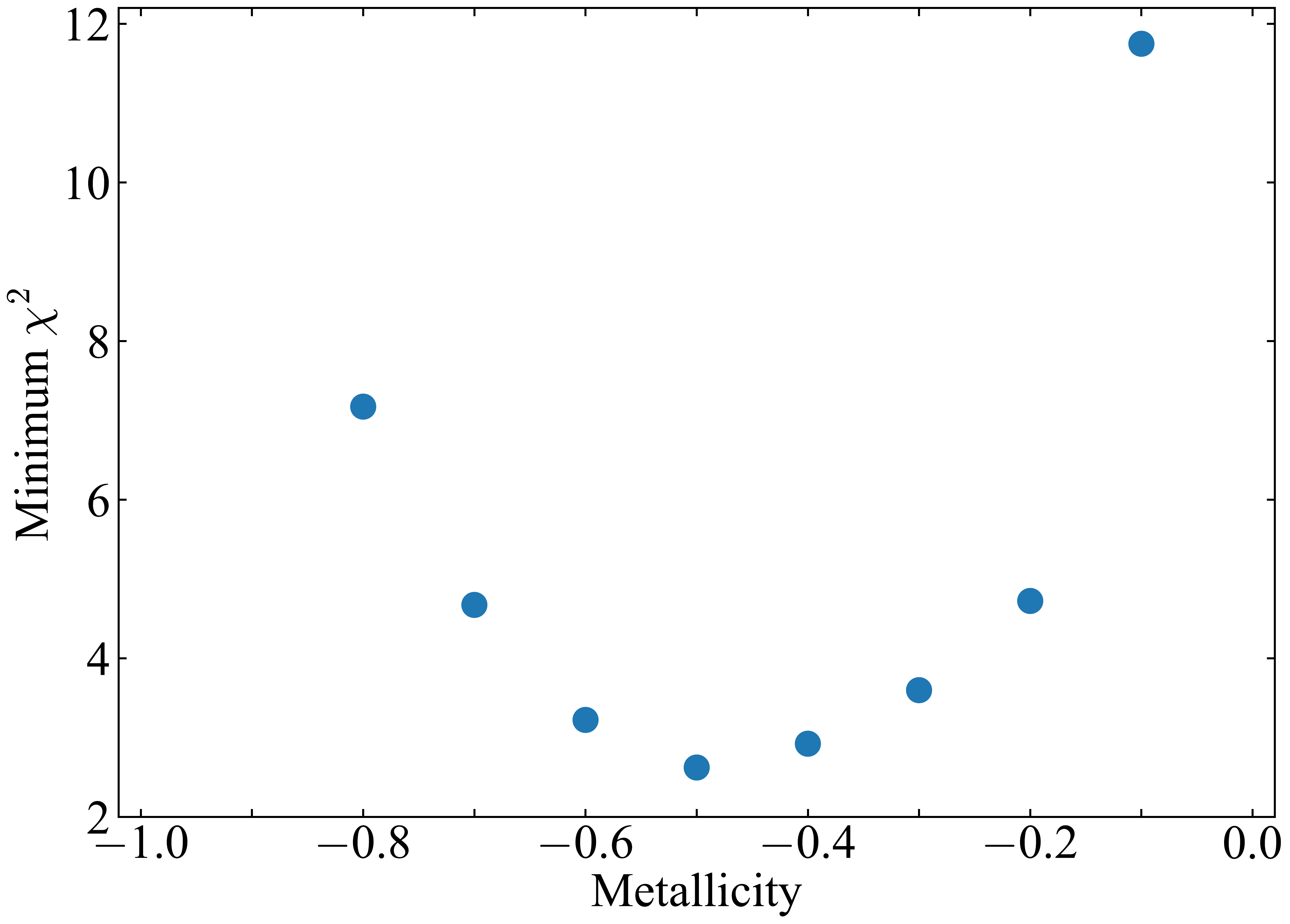}
\end{tabular}
\caption{Minimum $\chi^{2}$ as function of metallicity for the three-component dust model (see the text).}
 \label{themis_chi2}
\end{figure}

\begin{figure}
\centering
\begin{tabular}{c}
\includegraphics[width=0.99\hsize]{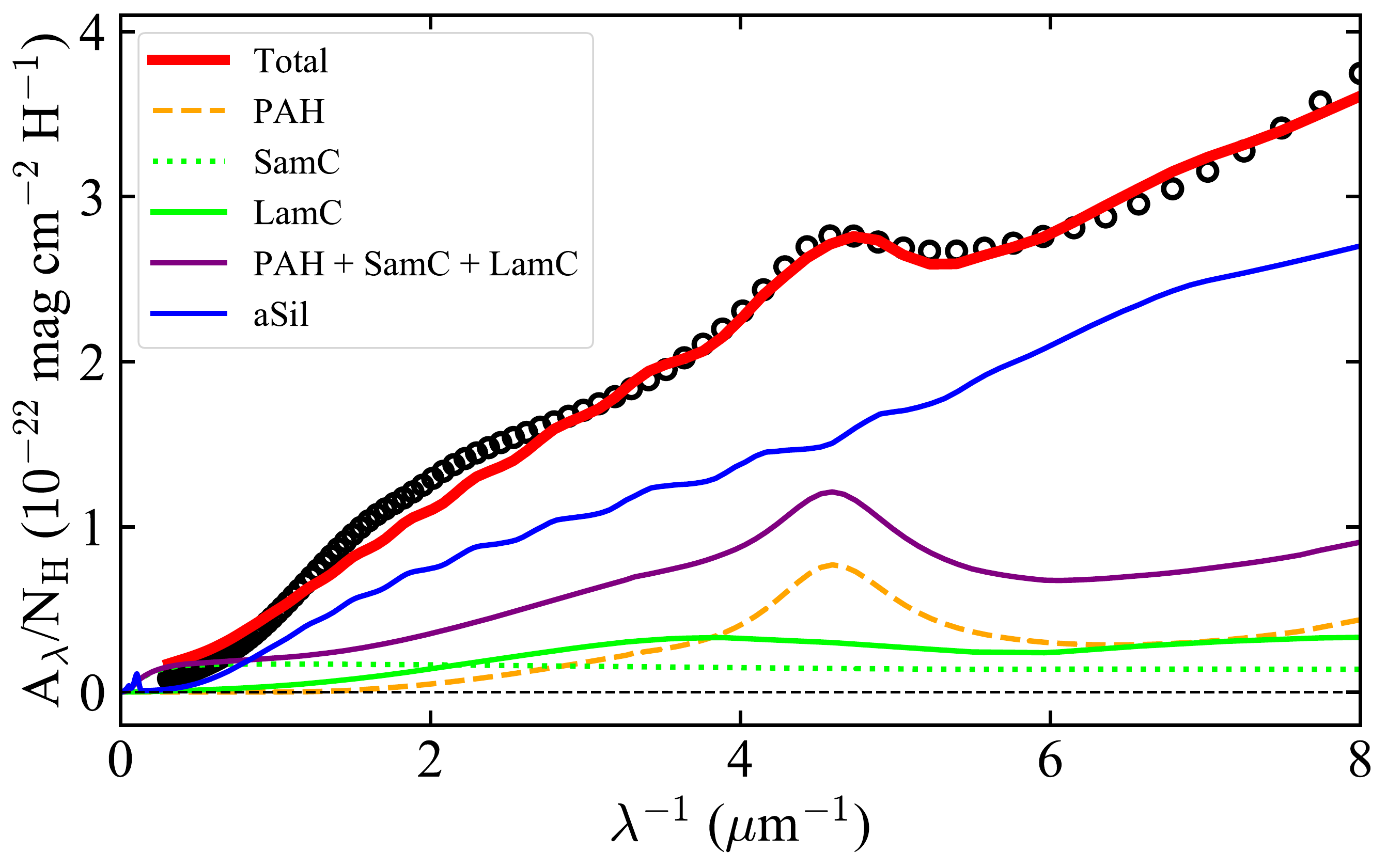}
\end{tabular}
\caption{Observed (black circles) and model (red line) extinction curves of GRB\,180325A for the three-component dust model. The other colored curves show the contributions from different dust components.}
 \label{themis_ext}
\end{figure}

\begin{table}
\caption{Parameters of the fit for our best dust model. The last column is the mass abundance per hydrogen for each dust component.} 
\centering 
 \setlength{\tabcolsep}{5.6pt}
\renewcommand{\arraystretch}{0.90}
\begin{tabular}{c c c c c c}
\hline\hline 
            &     $\sigma$   &     $a_{0}$      &       &        &       (M/M$_{\rm H}$)   \\
            &                     &     (nm)           &       &        &        \\
\hline
PAH      &     0.16          &     0.9             &       &        &       7.9\,$\times$\,10$^{-5}$   \\
SamC   &     1.26          &     10              &       &        &       8.3\,$\times$\,10$^{-5}$   \\
\hline 
                &     $\alpha$   &     $a_{\rm min}$      &   ($a_{\rm c}$, $a_{\rm t}$)    &   $\gamma$     &          \\
               &                     &     (nm)                     &                     (nm )                 &                        &        \\
\hline
LamC        &     $-$1.72    &       4.0                     &   (880, 70)        &   1.72     &     5.59\,$\times$\,10$^{-4}$   \\
aSil           &     $-$3.43    &       4.0                     &   (280, 120)    &   1.23     &     17.6\,$\times$\,10$^{-4}$   \\
\hline
\end{tabular}
\label{table3} 
\end{table}

\subsubsection{Three-Component Dust Model}

In this section, we try to model the GRB extinction curve using a model with three dust components: (i) PAHs, (ii) hydrogenated amorphous carbon (amC), and (iii) amorphous silicates (aSil). Here, the population of amorphous carbon dust is divided into small (SamC) and large (LamC) grains, but with continuous overall PAH\,+\,amC size distribution \citep[see figure\,1 in][]{2011A&A...525A.103C}. The PAH and SamC size distributions are assumed to be log-normal with $a_{0}$ and $\sigma$ representing the center radius and the width of the distribution, respectively. Moreover, the LamC and aSil size distributions have a power law form (i.e. $a^{\alpha}$) starting with the minimum size of $a_{\rm min}$ and an exponential cut-off of the following form

\begin{equation} \label{eq:13}
     \begin{cases}
      {\rm exp}(-[(a - a_{\rm t})/a_{\rm c}]^{\gamma}) & a \ge a_{\rm t} \\
      1 & a < a_{\rm t}\\
    \end{cases}       
\end{equation}

Here, our goal is to find the metallicity for which the model extinction curve best matches the observation. For this purpose, we first define a grid of metallicities ranging from $-$1.0 \citep[i.e. the lower limit reported by][]{2018ApJ...860L..21Z} to 0.0 with the step of 0.1\,dex. Then for each metallicity in the grid, we adjust the size distribution of the different dust populations until the minimum $\chi^{2}$ is reached. The $\chi^{2}$ is defined as in equation\,\ref{eq:9}. The model extinction, $A_{\rm mod}$, is calculated using the \texttt{DustEM} code which is freely available online\footnote{\url{https://www.ias.u-psud.fr/DUSTEM/dustem_code.html}}. Similar to section\,\ref{sect:two-comp}, we take $\sigma_{i}$\,=\,5\,$\sigma_{\rm obs}$ for $\lambda^{-1}$\,$<$\,2\,$\mu$m$^{-1}$ and $\sigma_{i}$\,=\,1\,$\sigma_{\rm obs}$ for 2\,$\mu$m$^{-1}$\,$<$\,$\lambda^{-1}$\,$<$\,8\,$\mu$m$^{-1}$ since the observed extinction in the infrared region is uncertain. Figure\,\ref{themis_chi2} shows the minimum $\chi^{2}$ value for each metallicity in the grid. The $\chi^{2}$ for $Z$\,=$-$1.0, $-$0.9, and $0.0$ are very large therefore, for the ease of illustration, they are not shown in Fig.\,\ref{themis_chi2}. As shown here, the best match between the model and the observed extinction curve occurs at $Z$\,=\,$-$0.5. Figure\,\ref{themis_ext} shows the best fit to the observed extinction curve, which is reached for $Z$\,=\,$-$0.5. Table\,\ref{table3} lists the parameters of the fit.

\section{Summary and Conclusion}

In this work, we presented a method to constrain the metallicities of high redshift (i.e. $z_{\rm abs}$\,$\ge$\,1.8) DLA absorbers using the observed extinction curves. We used the observed extinction curves of 13 quasars and one GRB in order to demonstrate the method. This is the first time the extinction curve is exploited to constrain the metallicities of extragalactic absorbers. We used the KK relation, which is a model-independent approach, to put some lower limits on the metallicities of the DLAs detected in the spectra of the quasars and the GRB. Here, the lower limits are calculated for three different dust compositions: 1) graphite and Fe-bearing silicates, 2) graphite, Fe-lacking silicates,  and pure iron, and 3) graphite, Fe-lacking silicates, and iron oxides. Interestingly, the results from the different dust compositions are consistent both with each other and with the observations.

The GRB extinction curve exhibits a very strong 2175\,\textup{\AA} extinction bump. We therefore tried to model the GRB extinction curve using dust models with two (i.e. graphite and silicates) and three (i.e. PAH, hydrogenated amorphous carbon, and silicates) dust components. The two-component model resulted in a metallicity of $Z\,\sim$\,$-$0.45 while the three-component model gives $Z\,\sim$\,$-$0.50. On the other hand, the lower limit from the KK approach for this DLA is $Z\,\ge$\,$-$0.60. Since the KK approach is model-independent, we propose $\sim$\,$-$0.6 as the new lower limit for the metallicity of the DLA detected along the line of sight to  GRB\,180325A. Modeling a large sample of extinction curves with 2175\,\textup{\AA} extinction bump and measured DLA metallicities would allow a thorough comparison between the KK and the model-dependent approach.

When the precise measurement of the metallicity of a DLA is not possible (e.g. due to the saturation of important absorption lines such as Zn\,{\sc ii}), the techniques employed in this paper can be used to constrain the metalicity, especially in absorbers associated with GRBs (due to their simple spectral energy distributions which follow a power law relation and lack prominent emission lines) and quasar absorbers with $z_{\rm abs}$\,$\sim$\,$z_{\rm em}$ such as eclipsing and ghostly DLAs. Since the presence of quasars broad emission lines could complicate the process of extracting the extinction curve from the quasar spectrum, these complications are minimum when $z_{\rm abs}$\,$\sim$\,$z_{\rm em}$, hence this technique could be better suited for such absorbers.

\section*{Acknowledgements}
The author would like to thank the referee for useful comments which improved the quality of the paper. The author would also like to thank Prof. Aigen Li and Dr. Tayyaba Zafar for useful discussions.

\bibliographystyle{aasjournal}
\bibliography{ref}

\begin{thebibliography}{}
\expandafter\ifx\csname natexlab\endcsname\relax\def\natexlab#1{#1}\fi
\providecommand{\url}[1]{\href{#1}{#1}}

\bibitem[{{Adamson} {et~al.}(1990){Adamson}, {Whittet}, \&
  {Duley}}]{1990MNRAS.243..400A}
{Adamson}, A.~J., {Whittet}, D.~C.~B., \& {Duley}, W.~W. 1990, \mnras, 243, 400

\bibitem[{{Allamandola} {et~al.}(1985){Allamandola}, {Tielens}, \&
  {Barker}}]{1985ApJ...290L..25A}
{Allamandola}, L.~J., {Tielens}, A.~G.~G.~M., \& {Barker}, J.~R. 1985, \apjl,
  290, L25

\bibitem[{{Anders} \& {Grevesse}(1989)}]{1989GeCoA..53..197A}
{Anders}, E., \& {Grevesse}, N. 1989, \gca, 53, 197

\bibitem[{{Asplund} {et~al.}(2009){Asplund}, {Grevesse}, {Sauval}, \&
  {Scott}}]{2009ARA&A..47..481A}
{Asplund}, M., {Grevesse}, N., {Sauval}, A.~J., \& {Scott}, P. 2009, \araa, 47,
  481

\bibitem[{{Bekki} {et~al.}(2015){Bekki}, {Hirashita}, \&
  {Tsujimoto}}]{2015ApJ...810...39B}
{Bekki}, K., {Hirashita}, H., \& {Tsujimoto}, T. 2015, \apj, 810, 39

\bibitem[{{Berg} {et~al.}(2015){Berg}, {Neeleman}, {Prochaska}, {Ellison}, \&
  {Wolfe}}]{2015PASP..127..167B}
{Berg}, T. A.~M., {Neeleman}, M., {Prochaska}, J.~X., {Ellison}, S.~L., \&
  {Wolfe}, A.~M. 2015, \pasp, 127, 167

\bibitem[{{Berg} {et~al.}(2016){Berg}, {Ellison}, {S{\'a}nchez-Ram{\'\i}rez},
  {Prochaska}, {Lopez}, {D'Odorico}, {Becker}, {Christensen}, {Cupani},
  {Denney}, \& {Worseck}}]{2016MNRAS.463.3021B}
{Berg}, T.~A.~M., {Ellison}, S.~L., {S{\'a}nchez-Ram{\'\i}rez}, R., {et~al.}
  2016, \mnras, 463, 3021

\bibitem[{{Compi{\`e}gne} {et~al.}(2011){Compi{\`e}gne}, {Verstraete}, {Jones},
  {Bernard}, {Boulanger}, {Flagey}, {Le Bourlot}, {Paradis}, \&
  {Ysard}}]{2011A&A...525A.103C}
{Compi{\`e}gne}, M., {Verstraete}, L., {Jones}, A., {et~al.} 2011, \aap, 525,
  A103

\bibitem[{{De Cia} {et~al.}(2013){De Cia}, {Ledoux}, {Savaglio}, {Schady}, \&
  {Vreeswijk}}]{2013A&A...560A..88D}
{De Cia}, A., {Ledoux}, C., {Savaglio}, S., {Schady}, P., \& {Vreeswijk}, P.~M.
  2013, \aap, 560, A88

\bibitem[{{Draine} \& {Lee}(1984)}]{1984ApJ...285...89D}
{Draine}, B.~T., \& {Lee}, H.~M. 1984, \apj, 285, 89

\bibitem[{{Draine} \& {Malhotra}(1993)}]{1993ApJ...414..632D}
{Draine}, B.~T., \& {Malhotra}, S. 1993, \apj, 414, 632

\bibitem[{{Dwek}(2016)}]{2016ApJ...825..136D}
{Dwek}, E. 2016, \apj, 825, 136

\bibitem[{{Fathivavsari}(2020{\natexlab{a}})}]{2020ApJ...888...85F}
{Fathivavsari}, H. 2020{\natexlab{a}}, \apj, 888, 85

\bibitem[{{Fathivavsari}(2020{\natexlab{b}})}]{2020ApJ...901..123F}
---. 2020{\natexlab{b}}, \apj, 901, 123

\bibitem[{{Fathivavsari} {et~al.}(2016){Fathivavsari}, {Petitjean},
  {Noterdaeme}, {P{\^a}ris}, {Finley}, {L{\'o}pez}, \&
  {Srianand}}]{2016MNRAS.461.1816F}
{Fathivavsari}, H., {Petitjean}, P., {Noterdaeme}, P., {et~al.} 2016, \mnras,
  461, 1816

\bibitem[{{Fathivavsari} {et~al.}(2015){Fathivavsari}, {Petitjean},
  {Noterdaeme}, {P{\^a}ris}, {Finley}, {L{\'o}pez}, {Srianand}, \&
  {S{\'a}nchez}}]{2015MNRAS.454..876F}
---. 2015, \mnras, 454, 876

\bibitem[{{Fathivavsari} {et~al.}(2017){Fathivavsari}, {Petitjean}, {Zou},
  {Noterdaeme}, {Ledoux}, {Kr{\"u}hler}, \& {Srianand}}]{2017MNRAS.466L..58F}
{Fathivavsari}, H., {Petitjean}, P., {Zou}, S., {et~al.} 2017, \mnras, 466, L58

\bibitem[{{Fathivavsari} {et~al.}(2018){Fathivavsari}, {Petitjean},
  {Jamialahmadi}, {Khosroshahi}, {Rahmani}, {Finley}, {Noterdaeme},
  {P{\^a}ris}, \& {Srianand}}]{2018MNRAS.477.5625F}
{Fathivavsari}, H., {Petitjean}, P., {Jamialahmadi}, N., {et~al.} 2018, \mnras,
  477, 5625

\bibitem[{{Fitzpatrick} \& {Massa}(1990)}]{1990ApJS...72..163F}
{Fitzpatrick}, E.~L., \& {Massa}, D. 1990, \apjs, 72, 163

\bibitem[{{Fitzpatrick} \& {Massa}(2007)}]{2007ApJ...663..320F}
---. 2007, \apj, 663, 320

\bibitem[{{Gao} {et~al.}(2020){Gao}, {Zhao}, {Gao}, {Jiang}, \&
  {Li}}]{2020P&SS..18304627G}
{Gao}, W., {Zhao}, R., {Gao}, J., {Jiang}, B., \& {Li}, J. 2020, \planss, 183,
  104627

\bibitem[{{Gordon} {et~al.}(2003){Gordon}, {Clayton}, {Misselt}, {Landolt}, \&
  {Wolff}}]{2003ApJ...594..279G}
{Gordon}, K.~D., {Clayton}, G.~C., {Misselt}, K.~A., {Landolt}, A.~U., \&
  {Wolff}, M.~J. 2003, \apj, 594, 279

\bibitem[{{Gudennavar} {et~al.}(2012){Gudennavar}, {Bubbly}, {Preethi}, \&
  {Murthy}}]{2012ApJS..199....8G}
{Gudennavar}, S.~B., {Bubbly}, S.~G., {Preethi}, K., \& {Murthy}, J. 2012,
  \apjs, 199, 8

\bibitem[{{Henning}(2010)}]{2010ARA&A..48...21H}
{Henning}, T. 2010, \araa, 48, 21

\bibitem[{{Jaeger} {et~al.}(1994){Jaeger}, {Mutschke}, {Begemann}, {Dorschner},
  \& {Henning}}]{1994A&A...292..641J}
{Jaeger}, C., {Mutschke}, H., {Begemann}, B., {Dorschner}, J., \& {Henning}, T.
  1994, \aap, 292, 641

\bibitem[{{Kemper} {et~al.}(2004){Kemper}, {Vriend}, \&
  {Tielens}}]{2004ApJ...609..826K}
{Kemper}, F., {Vriend}, W.~J., \& {Tielens}, A.~G.~G.~M. 2004, \apj, 609, 826

\bibitem[{{Kim} {et~al.}(1994){Kim}, {Martin}, \&
  {Hendry}}]{1994ApJ...422..164K}
{Kim}, S.-H., {Martin}, P.~G., \& {Hendry}, P.~D. 1994, \apj, 422, 164

\bibitem[{{Laor} \& {Draine}(1993)}]{1993ApJ...402..441L}
{Laor}, A., \& {Draine}, B.~T. 1993, \apj, 402, 441

\bibitem[{{Leger} \& {Puget}(1984)}]{1984A&A...137L...5L}
{Leger}, A., \& {Puget}, J.~L. 1984, \aap, 500, 279

\bibitem[{{Li}(2005)}]{2005ApJ...622..965L}
{Li}, A. 2005, \apj, 622, 965

\bibitem[{{Li} \& {Draine}(2001{\natexlab{a}})}]{2001ApJ...550L.213L}
{Li}, A., \& {Draine}, B.~T. 2001{\natexlab{a}}, \apjl, 550, L213

\bibitem[{{Li} \& {Draine}(2001{\natexlab{b}})}]{2001ApJ...554..778L}
---. 2001{\natexlab{b}}, \apj, 554, 778

\bibitem[{{Li} {et~al.}(2008){Li}, {Liang}, {Kann}, {Wei}, {Klose}, \&
  {Wang}}]{2008ApJ...685.1046L}
{Li}, A., {Liang}, S.~L., {Kann}, D.~A., {et~al.} 2008, \apj, 685, 1046

\bibitem[{{Liang} \& {Li}(2009)}]{2009ApJ...690L..56L}
{Liang}, S.~L., \& {Li}, A. 2009, \apjl, 690, L56

\bibitem[{{Liang} \& {Li}(2010)}]{2010ApJ...710..648L}
---. 2010, \apj, 710, 648

\bibitem[{{Ma} {et~al.}(2020){Ma}, {Zhu}, {Yan}, {You}, \&
  {Su}}]{2020MNRAS.497.2190M}
{Ma}, X.-Y., {Zhu}, Y.-Y., {Yan}, Q.-B., {You}, J.-Y., \& {Su}, G. 2020,
  \mnras, 497, 2190

\bibitem[{{Mathis} {et~al.}(1977){Mathis}, {Rumpl}, \&
  {Nordsieck}}]{1977ApJ...217..425M}
{Mathis}, J.~S., {Rumpl}, W., \& {Nordsieck}, K.~H. 1977, \apj, 217, 425

\bibitem[{{Mishra} \& {Li}(2015)}]{2015ApJ...809..120M}
{Mishra}, A., \& {Li}, A. 2015, \apj, 809, 120

\bibitem[{{Mishra} \& {Li}(2017)}]{2017ApJ...850..138M}
---. 2017, \apj, 850, 138

\bibitem[{{Mulas} {et~al.}(2013){Mulas}, {Zonca}, {Casu}, \&
  {Cecchi-Pestellini}}]{2013ApJS..207....7M}
{Mulas}, G., {Zonca}, A., {Casu}, S., \& {Cecchi-Pestellini}, C. 2013, \apjs,
  207, 7

\bibitem[{{Papoular} \& {Papoular}(2009)}]{2009MNRAS.394.2175P}
{Papoular}, R.~J., \& {Papoular}, R. 2009, \mnras, 394, 2175

\bibitem[{{Pendleton} \& {Allamandola}(2002)}]{2002ApJS..138...75P}
{Pendleton}, Y.~J., \& {Allamandola}, L.~J. 2002, \apjs, 138, 75

\bibitem[{{Poteet} {et~al.}(2015){Poteet}, {Whittet}, \&
  {Draine}}]{2015ApJ...801..110P}
{Poteet}, C.~A., {Whittet}, D. C.~B., \& {Draine}, B.~T. 2015, \apj, 801, 110

\bibitem[{{Press} {et~al.}(1992){Press}, {Teukolsky}, {Vetterling}, \&
  {Flannery}}]{1992nrfa.book.....P}
{Press}, W.~H., {Teukolsky}, S.~A., {Vetterling}, W.~T., \& {Flannery}, B.~P.
  1992, {Numerical recipes in FORTRAN. The art of scientific computing}

\bibitem[{{Purcell}(1969)}]{1969ApJ...158..433P}
{Purcell}, E.~M. 1969, \apj, 158, 433

\bibitem[{{Savage} {et~al.}(1992){Savage}, {Cardelli}, \&
  {Sofia}}]{1992ApJ...401..706S}
{Savage}, B.~D., {Cardelli}, J.~A., \& {Sofia}, U.~J. 1992, \apj, 401, 706

\bibitem[{{Selsing} {et~al.}(2016){Selsing}, {Fynbo}, {Christensen}, \&
  {Krogager}}]{2016A&A...585A..87S}
{Selsing}, J., {Fynbo}, J.~P.~U., {Christensen}, L., \& {Krogager}, J.-K. 2016,
  \aap, 585, A87

\bibitem[{{Sofia} {et~al.}(1994){Sofia}, {Cardelli}, \&
  {Savage}}]{1994ApJ...430..650S}
{Sofia}, U.~J., {Cardelli}, J.~A., \& {Savage}, B.~D. 1994, \apj, 430, 650

\bibitem[{{Spitzer} \& {Jenkins}(1975)}]{1975ARA&A..13..133S}
{Spitzer}, L., J., \& {Jenkins}, E.~B. 1975, \araa, 13, 133

\bibitem[{{Spitzer} \& {Fitzpatrick}(1995)}]{1995ApJ...445..196S}
{Spitzer}, Lyman, J., \& {Fitzpatrick}, E.~L. 1995, \apj, 445, 196

\bibitem[{{Stecher} \& {Donn}(1965)}]{1965ApJ...142.1681S}
{Stecher}, T.~P., \& {Donn}, B. 1965, \apj, 142, 1681

\bibitem[{{Steglich} {et~al.}(2010){Steglich}, {J{\"a}ger}, {Rouill{\'e}},
  {Huisken}, {Mutschke}, \& {Henning}}]{2010ApJ...712L..16S}
{Steglich}, M., {J{\"a}ger}, C., {Rouill{\'e}}, G., {et~al.} 2010, \apjl, 712,
  L16

\bibitem[{{Vladilo}(1998)}]{1998ApJ...493..583V}
{Vladilo}, G. 1998, \apj, 493, 583

\bibitem[{{Voshchinnikov} \& {Henning}(2010)}]{2010A&A...517A..45V}
{Voshchinnikov}, N.~V., \& {Henning}, T. 2010, \aap, 517, A45

\bibitem[{{Wang} {et~al.}(2015){Wang}, {Li}, \& {Jiang}}]{2015ApJ...811...38W}
{Wang}, S., {Li}, A., \& {Jiang}, B.~W. 2015, \apj, 811, 38

\bibitem[{{Weingartner} \& {Draine}(2001)}]{2001ApJ...548..296W}
{Weingartner}, J.~C., \& {Draine}, B.~T. 2001, \apj, 548, 296

\bibitem[{{Welty} {et~al.}(1997){Welty}, {Lauroesch}, {Blades}, {Hobbs}, \&
  {York}}]{1997ApJ...489..672W}
{Welty}, D.~E., {Lauroesch}, J.~T., {Blades}, J.~C., {Hobbs}, L.~M., \& {York},
  D.~G. 1997, \apj, 489, 672

\bibitem[{{Xiang} {et~al.}(2011){Xiang}, {Li}, \&
  {Zhong}}]{2011ApJ...733...91X}
{Xiang}, F.~Y., {Li}, A., \& {Zhong}, J.~X. 2011, \apj, 733, 91

\bibitem[{{Zafar} {et~al.}(2011){Zafar}, {Watson}, {Fynbo}, {Malesani},
  {Jakobsson}, \& {de Ugarte Postigo}}]{2011A&A...532A.143Z}
{Zafar}, T., {Watson}, D., {Fynbo}, J.~P.~U., {et~al.} 2011, \aap, 532, A143

\bibitem[{{Zafar} {et~al.}(2012){Zafar}, {Watson}, {El{\'\i}asd{\'o}ttir},
  {Fynbo}, {Kr{\"u}hler}, {Schady}, {Leloudas}, {Jakobsson}, {Th{\"o}ne},
  {Perley}, {Morgan}, {Bloom}, \& {Greiner}}]{2012ApJ...753...82Z}
{Zafar}, T., {Watson}, D., {El{\'\i}asd{\'o}ttir}, {\'A}., {et~al.} 2012, \apj,
  753, 82

\bibitem[{{Zafar} {et~al.}(2018{\natexlab{a}}){Zafar}, {Heintz}, {Fynbo},
  {Malesani}, {Bolmer}, {Ledoux}, {Arabsalmani}, {Kaper}, {Campana},
  {Starling}, {Selsing}, {Kann}, {de Ugarte Postigo}, {Schweyer},
  {Christensen}, {M{\o}ller}, {Japelj}, {Perley}, {Tanvir}, {D'Avanzo},
  {Hartmann}, {Hjorth}, {Covino}, {Sbarufatti}, {Jakobsson}, {Izzo},
  {Salvaterra}, {D'Elia}, \& {Xu}}]{2018ApJ...860L..21Z}
{Zafar}, T., {Heintz}, K.~E., {Fynbo}, J.~P.~U., {et~al.} 2018{\natexlab{a}},
  \apjl, 860, L21

\bibitem[{{Zafar} {et~al.}(2018{\natexlab{b}}){Zafar}, {Watson}, {M{\o}ller},
  {Selsing}, {Fynbo}, {Schady}, {Wiersema}, {Levan}, {Heintz}, {de Ugarte
  Postigo}, {D'Elia}, {Jakobsson}, {Bolmer}, {Japelj}, {Covino}, {Gomboc}, \&
  {Cano}}]{2018MNRAS.479.1542Z}
{Zafar}, T., {Watson}, D., {M{\o}ller}, P., {et~al.} 2018{\natexlab{b}},
  \mnras, 479, 1542

\bibitem[{{Zonca} {et~al.}(2011){Zonca}, {Cecchi-Pestellini}, {Mulas}, \&
  {Malloci}}]{2011MNRAS.410.1932Z}
{Zonca}, A., {Cecchi-Pestellini}, C., {Mulas}, G., \& {Malloci}, G. 2011,
  \mnras, 410, 1932

\bibitem[{{Zuo} {et~al.}(2020){Zuo}, {Li}, \& {Zhao}}]{2020arXiv201109440Z}
{Zuo}, W.~B., {Li}, A., \& {Zhao}, G. 2020, arXiv e-prints, arXiv:2011.09440

\end{thebibliography}

\end{document}